\documentclass[twocolumn,showpacs,preprintnumbers,amsmath,amssymb, pre]{revtex4}
\usepackage{graphicx}% Include figure files
\usepackage{dcolumn}% Align table columns on decimal point
\usepackage{bm}% bold math

\begin{document}

%\preprint{APS/123-QED}

\title{How does torsional rigidity affect the wrapping transition of a semiflexible chain around a spherical core?}% Force line breaks with \\

\author{Yuji Higuchi}
%\email[]{}
%\homepage[]{Your web page}
%\thanks{}
%\altaffiliation{}
\affiliation{Department of Physics, Graduate School of Science, Kyoto University, Kyoto 606-8502, Japan}

\author{Takahiro Sakaue}
%\email[]{}
%\homepage[]{Your web page}
%\thanks{}
%\altaffiliation{}
\affiliation{Department of Physics, Graduate School of Science, Kyushu University and PRESTO, Japan Science and Technology Agency (JST), 4-1-8 Honcho Kawaguchi, Saitama 332-0012, Japan}

\author{Kenichi Yoshikawa}
\email[Electronic address:]{yoshikaw@scphys.kyoto-u.ac.jp}
%\homepage[]{Your web page}
%\thanks{}
%\altaffiliation{}
\affiliation{Department of Physics, Graduate School of Science, Kyoto University and Spatio-Temporal Order Project, ICORP, JST, Kyoto 606-8502, Japan}

\date{\today}% It is always \today, today,
             %  but any date may be explicitly specified

\begin{abstract}
We investigated the effect of torsional rigidity of a semiflexible chain on the wrapping transition around a spherical core, as a model of nucleosome, the fundamental unit of chromatin. Through molecular dynamics simulation, we show that the torsional effect has a crucial effect on the chain wrapping around the core under the topological constraints. In particular, the torsional stress (i) induces the wrapping/unwrapping transition, and (ii) leads to a unique complex structure with an antagonistic wrapping direction which never appears without the topological constraints. We further examine the effect of the stretching stress for the nucleosome model, in relation to the unique characteristic effect of the torsional stress on the manner of wrapping.
\end{abstract}

%\pacs{05.65.+b, 68.03.Cd, 82.70.Gg, 89.75.Kd}% PACS, the Physics and Astronomy
                             % Classification Scheme.
%\keywords{Suggested keywords}%Use showkeys class option if keyword
                              %display desired
\maketitle

\section{Introduction}
The structure of DNA wrapping a cationic protein is called nucleosome, which is a basic unit of chromatin in living cells~\cite{Wolffe}. In physics, a spherical or cylindrical core and a semiflexible chain are adopted as a model to feature nucleosome structure~\cite{Cond_Schiessel,SCDB_Langowski,PRL_Sakaue,PRE_Yanao,PRE_Li,PRE_Li2,PRL_Kunze,PRE_Kunze}. In contrast to a flexible chain, a semiflexible chain wraps the core orderly~\cite{PRE_Sakaue}. Under usual aqueous conditions, the persistence length $l_p\simeq 50$ nm of DNA is large compared to its molecular thickness $d\simeq 2$ nm. Therefore DNAs with contour length $L\gg 1$ $\mu$m is characterized as a typical semiflexible chain: for example, the contour length of eukaryote DNA is longer than 500 $\mu$m. DNA wraps a histone core, whose diameter is 11 nm, 1.75 times orderly. To reveal structures and mechanisms of nucleosome, some groups study by pulling nucleosome~\cite{PRE_Sakaue,PNAS_Brower,PRL_Kulic,Nature_Bennink,PNAS_Mihardja,JPCB_Wocjan}. It was found that unwrapping is generated in a stepwise manner, i.e., the chain unwraps the core twice to once, then once to zero when the chain is pulled. These wrapping transitions are controlled by several factors, an interaction between them, the chain stiffness, and the sphere size. Recently, it is revealed that the other factor is also important to nucleosome structures and mechanisms. Yanao {\it et al}.~\cite{PRE_Yanao} shows that DNA wraps protein core left-handed-like considering  the coupling between bending and twisting rigidities owe to the chirality of the right-handed double helix structure of DNA. In the system of a chain wrapping a core, topology is one of the important problems to reveal the mechanism and the structure of nucleosome.

In prokaryote cell, DNA is a circular chain and topologically constrained. In these cases, topological number $Lk=Wr+Tw$ is conserved~\cite{Biopolymers_Klenin}. $Wr$ is calculated through the chain conformation. $Tw$ is total twisting of the chain. The effects of topological constraints and torsional rigidity on the conformation have been studied in circular polymer~\cite{Biopolymers_Chirico,PRE_Marko,Biophys_Chirico,CPC_Velichko}. 
In eukaryote cell, DNA is compactly stored by forming a hierarchical structure. In the lowest level, nucleosomes are packed into the 30 nm chromatin fiber. In interphase cell, this fiber is further organized into Lampbrush chromosomes, which consists of highly condensed chromatin parts and extended chromatin loop parts emanating from condensed parts. Most DNA parts are generally not expressed, which are highly condensed. On the other hand, the chromatin loops are expressed actively~\cite{Cell}. The chromatin loop is fixed both ends and topologically constrained. In eukaryote, the structure of DNA is also influenced by the torsional rigidity. There are some researches to study how the topological constraints influence the conformation of nucleosomes or chromatin fiber~\cite{PNAS_Crick,PRE_Barbi,FEBS_Mozziconacci}. However, there is an unsolved question how torsional stress under topological constraints affects the wrapping manner of the nucleosome. In this article, we will study the relation between torsional effect and the stability of the wrapping manner of nucleosome. We use a coarse-grained torsional model~\cite{Biopolymers_Chirico} to reveal the simple relation. We use a semiflexible chain and a spherical core model inspired by nucleosome. 

We report the nontrivial relation between torsional effect and the wrapping transition of semiflexible polymer based on a systematic Molecular Dynamics simulation. Our paper is organized as follows: In Sec.~I\hspace{-.1em}I, we introduce the model of a core and a semifexible chain which is adapted in the simulation. In Sec.~I\hspace{-.1em}I\hspace{-.1em}I, we show our simulational results. In Sec.~I\hspace{-.1em}V, we discuss the wrapping transition in terms of free energy. Further results are interpreted in terms of free energy. Finally we conclude in Sec.~V.
\section{Methods}
To investigate the wrapping and unwrapping transition of semiflexible polymers, we carried out off-lattice Molecular Dynamics simulations in three-dimensional space. The model of one polymer and one core is essentially the same as that studied previously in Ref.~\cite{PRL_Sakaue,PRE_Sakaue}. Torsional potential is introduced following methods proposed in Ref.~\cite{Biopolymers_Chirico}. The polymer is described as a beads-spring with positions ${\bf r}_i$; bond vectors ${\bf u}_i=({\bf r}_{i+1}-{\bf r}_i)/|{\bf r}_{i+1}-{\bf r}_i|$. The position of a spherical core is denoted by ${\bf r}_c$. To account for the material twisting, two normal vectors ${\bf f}_i$ and ${\bf v}_i$ are set at the gravity center of monomers ${\bf r}_i$ and obey ${\bf v}_i={\bf u}_i \times {\bf f}_i$ and ${\bf f}_i\cdot {\bf u}_i=0$. The potential energy of the system is represented by the following five terms;
\begin{equation}
\frac{U_{bond}}{k_BT}=\sum_{i=1}^{N} \frac{k_{bond}}{2}(| {\bf r}_{i+1}-{\bf r}_i| -\sigma)^2
\label{eq1}
\end{equation}
\begin{equation}
\frac{U_{bend}}{k_BT}=\sum_{i=1}^{N} \frac{k_{\theta}}{2}(1-\cos\theta _i)^2 
\end{equation}
\begin{equation}
\frac{U_{LJ}}{k_BT}=\epsilon\sum_{| i-j| >1}((\frac{\sigma}{| {\bf r}_i-{\bf r}_j|})^{12}-2(\frac{\sigma}{| {\bf r}_i-{\bf r}_j|})^6)
\end{equation}
\begin{equation}
\frac{U_{torsion}}{k_BT}=\sum_{i=1}^{N}\frac{k_{tor}}{2}(\alpha_i+\gamma_i)^2
\end{equation}
\begin{equation}
\frac{U_{LJhistone}}{k_BT}=\epsilon_{histone}\sum((\frac{\sigma \prime}{| {\bf r}_i-{\bf r}_c|})^{12}-2(\frac{\sigma \prime}{| {\bf r}_i-{\bf r}_c|})^6)
\label{eq5}
\end{equation}
where $\theta _i$ is the angle between adjacent bond vectors, $\alpha _i +\gamma _i$ is twist angle, which satisfies $\sin(\alpha _i+\gamma _i)=({\bf v}_i\cdot {\bf f}_{i+1}-{\bf f}_i\cdot {\bf v}_{i+1})/(1+{\bf u}_i\cdot {\bf u}_{i+1})$ and $\cos(\alpha _i+\gamma _i)=({\bf f}_i\cdot {\bf f}_{i+1}+{\bf v}_i\cdot {\bf v}_{i+1})/(1+{\bf u}_i\cdot {\bf u}_{i+1})$. The monomer size $\sigma$ and $k_BT$ are chosen as the unit length and energy, respectively.
We set the spring constant $k_{bond}=500$, the bending elasticity $k_{\theta } =60$, which corresponds to a persistence length $l_p\simeq 10\sigma$, and torsional energy $k_{tor}=60$. The excluded-volume effect is included in the Lennard-Jones potential $U_{LJ}$, we set $\epsilon =0.20$. For attractive energy and excluded-volume effect between a polymer and a core, we set $\epsilon_{histone}=5.0$. The size of core $\sigma_{histone}=2.0\sigma$ and $\sigma \prime =(\sigma_{histone}+\sigma )/2$. We set $N=60$ polymer beads and one core. In this article, $\sigma$, $l_p$, and $\sigma_{histone}$ correspond to 5, 50, and 10 nm, which are good approximations to DNA and histone core (see introduction).

Even though a tightly wrapped complex is formed, the core slides along the chain and prefers positioning at the chain end~\cite{PRL_Sakaue}. In order to eliminate possible end effects of the chain and to get clear-cut configuration, we fix the center of the core as like Ref.~\cite{PRE_Sakaue}.

The force ${\bf f}$ and the torque $\tau$ is calculated from Eq.~(1) - (5) and Eq.~(4), respectively. The monomers obey the stochastic dynamics described by Langevin equation without momentum term
\begin{equation}
\eta \frac{d{\bf r}_i}{dt}={\bf f}_i^U+\bm{\xi}_i
\end{equation}
\begin{equation}
\eta_R\frac{d\Psi_i}{dt}=\psi_i^U+\xi_i^{\psi}
\end{equation}
where $\Psi_i$ is the Euler angle of pure rotation, $\eta$ is the drag coefficient for drift, and $\eta_R$ is the drag coefficient for rotation. The constant $\tau=\eta \sigma ^2/k_BT$ is chosen to be the unit for the time scale. We set the time step as $dt=2.5\times 10^{-5}\tau$. The Brownian force $\bm{\xi}_i$ and torque $\xi_i^{\psi}$ satisfy the fluctuation dissipation theorem
\begin{equation}
<\bm{\xi}_i(t)\bm{\xi}_i(t\prime)>=6k_BT\eta\delta_{ij}\delta(t-t\prime)
\end{equation}
\begin{equation}
<\xi_i^{\tau}(t)\xi_j^{\tau}(t\prime)>=2k_BT\eta_R\delta_{ij}\delta(t-t\prime)
\end{equation}
We set $\eta=1.0$ and $\eta_R=0.213$~\cite{Doi}.

Let us introduce the following order parameter
\begin{equation}
P=\sum_{i=1}^{N} \rho (i)
\end{equation}
where $\rho(i)$ is an indicator of the pair contact:
$\rho(i)=1$ if the number of monomers, which satisfies $| {\bf r}_i-{\bf r}_c| <r_l$ and otherwise $\rho_{i,j}=0 $. In the following discussion we set $r_l=2.5\sigma$.
This quantity represents the degree of contact between the polymer and the core. As we shall see later, $P$ is directly related to the wrapping number $N_W$ which measures how many times the chain wraps the core.

In this paper, we study the chain conformation under three different manner of constraints; (i) twist one end (the other end is fixed), (ii) fixed both ends, and (iii) pull one end (the other end is fixed). In these handlings, the position of five monomers at the end (${\bf r}_{N-4}$ to ${\bf r}_{N}$) are fixed and $\Psi$ of those are also fixed. The positions of the other end monomers (${\bf r}_1$ to ${\bf r}_5$) are fixed in the case of (i) and (ii). In the case of (iii), those are fixed to $y$ and $z$ directions and pulled to $x$ direction. $\Psi_1$ to $\Psi_5$ are twisted in (i) and fixed in (ii) and (iii). We set ${\bf r}_N=(60,0,0)$, ${\bf r}_{N-1}=(59,0,0)$, ${\bf r}_{N-2}=(58,0,0)$, ${\bf r}_{N-3}=(57,0,0)$, and ${\bf r}_{N-4}=(56,0,0)$. The monomers of ${\bf r}_1$ to ${\bf r}_5$ are set at $(y,z)=(0,0)$. The core is set at ${\bf r}_c=(45,2,0)$. We manipulate 5 monomers of both ends for the reason to prevent the chain crossing over the chain end and for convenience to twist and fix.

We introduce topological number $Lk$, $Wr$, and $Tw$~\cite{Biopolymers_Klenin}. In this paper, both ends of the chain are fixed, $Lk$ is conserved in the case of (ii) and (iii).  In the case of (i), we decrease or increase $Lk$ linearly. $Lk$ is defined as follows
\begin{equation}
Lk=Wr+Tw
\end{equation}
$Wr$ is defined as follows
\begin{equation}
4\pi Wr=\int_C \int_C (d{\bf r}_1\times d{\bf r}_2)\cdot{\bf r}_{1,2}/| {\bf r}_{1,2}| ^3
\end{equation}
where ${\bf r}_1$ and ${\bf r}_2$ are the points passing along the closed curve $C$, ${\bf r}_{12}={\bf r}_2-{\bf r}_1$. In the calculation, we use Ref.~\cite{Biopolymers_Klenin}.
$Tw$ is defined as follows
\begin{equation}
2\pi Tw=\sum_{i=1}^{N}(\alpha_i+\gamma_i)
\end{equation}
We checked $Lk$ every 100 steps in order to look out the unphysical topological breaking.
\section{results}
\begin{figure}
\includegraphics{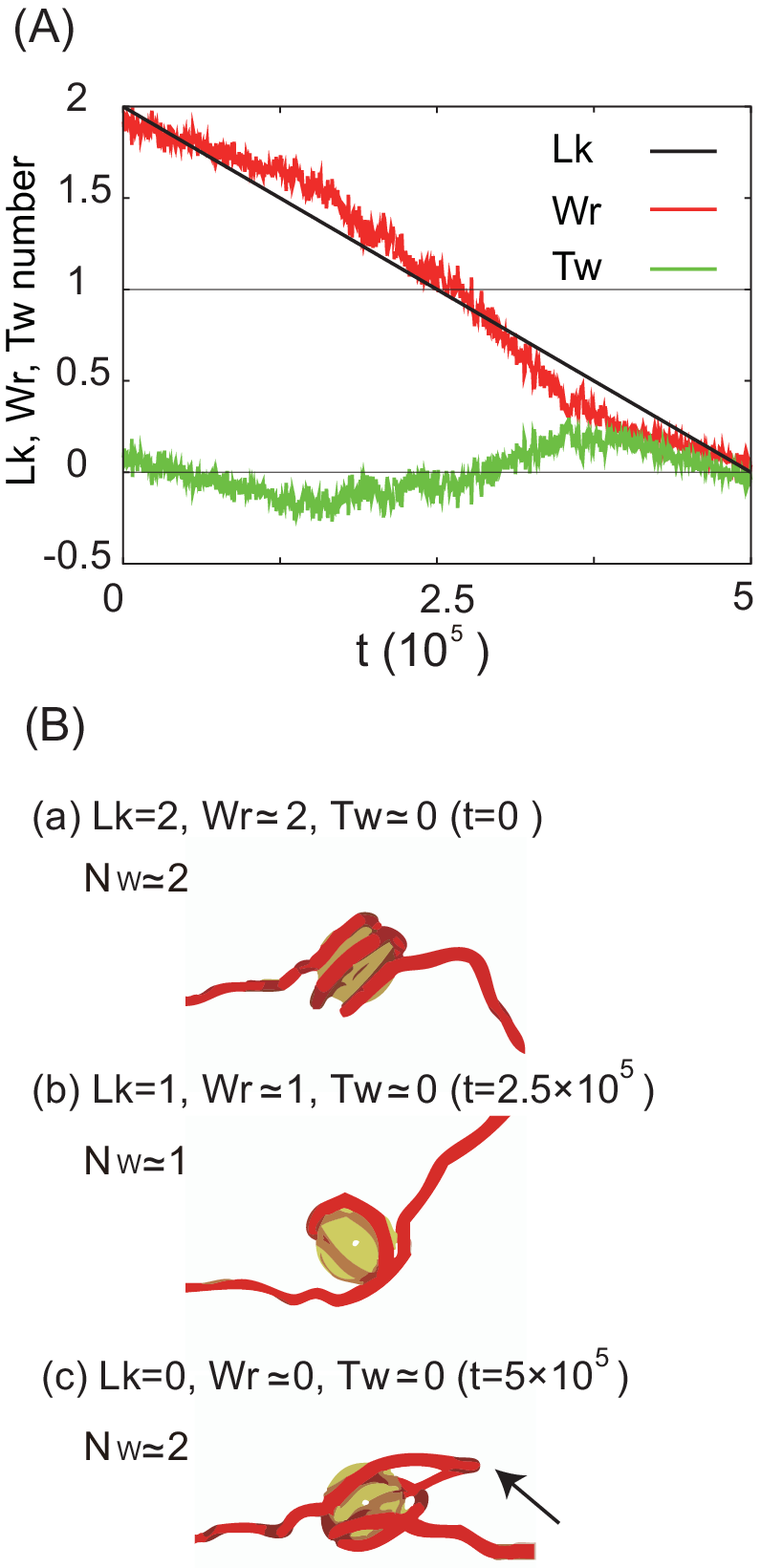}
\caption{Quasi-static time evolution of the topological numbers and the chain conformation. (A) Changes of $Wr$ (red) and $Tw$ (green) numbers accompanied by linear decrease of $Lk=2$ to 0 versus time steps. The data is an average of ten processes. (B) Typical snapshots of the chain and the core: (a) $Lk=2$ (b) $Lk=1$ (c) $Lk=0$.}
\label{Fig2}
\end{figure}

\begin{figure}
\includegraphics{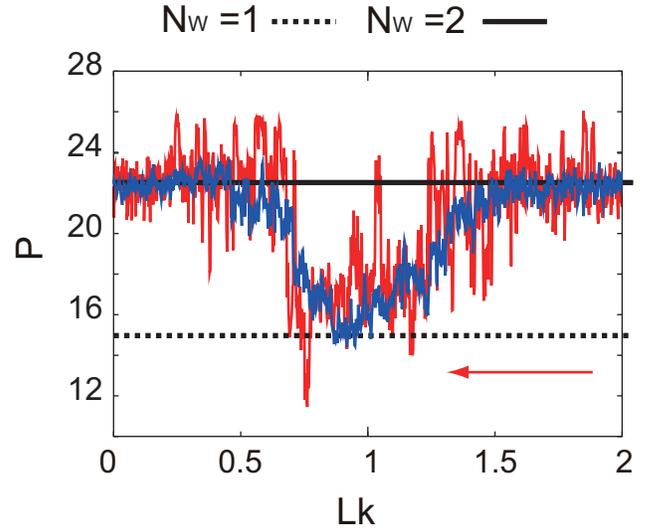}
\caption{The wrapping number $N_W$ versus $Lk$; $N_W=2$ (P=22.5) and $N_W=15$ are described by solid and dotted lines respectively. The chain is twisted from $Lk=2$ to 0 linearly with time (red line). The blue line is an average of ten processes.}
\label{Fig4}
\end{figure}

We first study how the chain wraps the core with twisting; $Lk$ (control parameter) is decreased linearly from $Lk=2$ to $Lk= 0$ in $2\times10^9$ steps ($t=5\times 10^4\tau$) without topological breaking. The end to end distance is 33.5$\sigma$. Fig.~\ref{Fig2} shows (A) evolution of topological numbers Lk, Wr, and Tw versus time and (B) snapshots of typical complex structures at $t=0$, $2.5\times10^4$, and $5\times10^4$, respectively. The initial linking number Lk=2 indicates that the chain accommodates itself to the state with no twisting penalty by wrapping around the core twice in a left-handed fashion ($Wr=2$) (Fig.~\ref{Fig2} (B) (a)). When the linking number is decreased, the complex first responds by the negative twisting while the writhing number kept almost fixed. This linear response regime is terminated at $Lk\simeq 1.5$, around which the complex exhibits the global structural change, i.e., unwrapping transition from $Wr\simeq 2$ to $Wr\simeq 1$, and releases the torsional stress. Because of the smallness of the system, the transition is not very sharp, but there are a finite range of the control parameter ($Lk$), where the bimodal distribution is realized. The state with $Wr\simeq 1$ and $Tw\simeq 0$ is stable around $Lk=1$ (Fig.~\ref{Fig2} (B) (b)). By further decreasing Lk, the complex exhibits the second global structural transition at $Lk\simeq 0.5$ to the state with $Wr\simeq 0$. This, however, does not correspond to the unwrapping, but rather to the wrapping transition. Fig.~\ref{Fig4} plots the order parameter $P$ as a function of $Lk$, which clearly demonstrates the second structural transition as a wrapping. The inspection of the snapshot indicates that each of two turns is characterized by the opposite handeness, and this antagonistic wrapping results in $Wr=0$. To realize such an antagonistic wrapping, there must be a "loop" in which the chain segment cannot be attached to the core (designated by an arrow in Fig.~\ref{Fig2} (B) (c)). We have confirmed that its energetic stability is almost the same as that of the natural wrapped state; the total energy and the elastic bending energy of the chain in the case of $Lk=0$ are almost same as (little less than) the energy in the case of $Lk=2$. The mean total internal energy $u=U/(N-10)$ ($U$ is sumation of eq.~(\ref{eq1}) to (\ref{eq5})), the mean elastic bending energy $u_{bend}=U_{bend}/(N-10)$, and the mean adsorption energy $u_{ad}=U_{ad}/(N-10)$ are given as $u=1.264$, $u_{bend}=1.066$, $u_{ad}=-0.981$ in the case of $Lk=0$ and $u=1.269$, $u_{bend}=1.086$, $u_{ad}=-1.024$ in the case of $Lk=2$. It is interesting to note that although the presence of such a loop is an unnatural form, i.e., never observed in the system without topological constraint, the fluctuation of $P$ is more suppressed in the antagonistic wrapping state ($N_W=2$ and $Wr=0$) than in the regular wrapping state ($N_W=2$ and $Wr=2$). This structural stability of the complex is probably rendered by the steric hinderance of the chain segments which cannot cross one another (topological effect of the chain).

\begin{figure}
\includegraphics{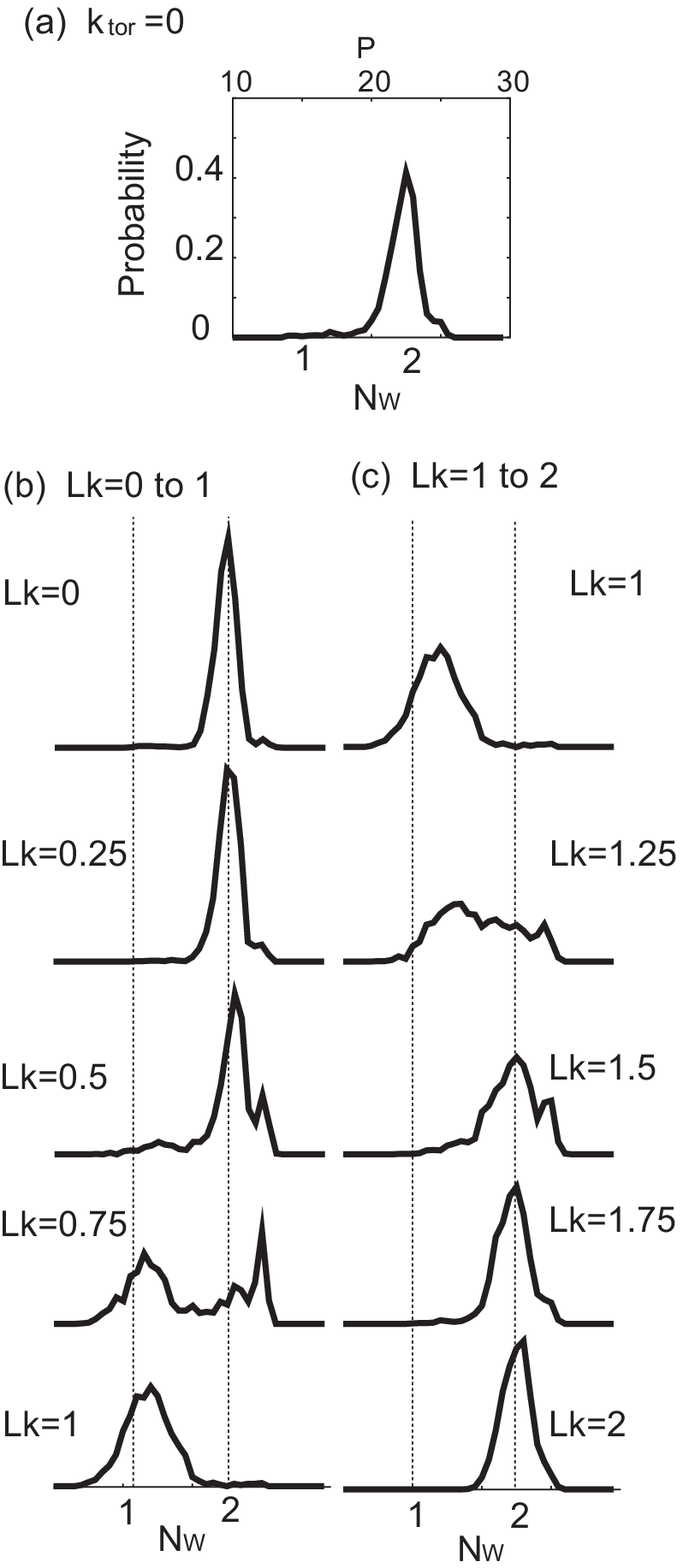}
\caption{The distribution of the wrapping number with $Lk$: (a) Without torsional stress. (b) At $Lk=0$, 0.25, 0.5, 0.75, and 1. (c) At $Lk=1$, 1.25, 1.5, 1.75, and 2.}
\label{Fig5}
\end{figure}

We study how the chain wraps the core with $Lk$. The both ends of the chain are fixed. The end to end distance is $33.5 \sigma$. Fig.~\ref{Fig5} shows the distribution of wrapping number versus $Lk$.
In $k_{tor} =0$ (without torsional effect), the peak is at $N_W=2$, which indicates that the chain wraps the core twice. In $Lk=0$, $Lk=0.25$, and $Lk=0.5$, the peak is at $N_W=2$. These results are alomost same as the result of Fig.~\ref{Fig4}. In $Lk=0.75$, there are two peaks at $N_W=1$ and $N_W>2$, which indicates that the chain wraps the core once and more than twice. In $Lk\simeq 0.75$, the chain wraps and unwraps the core (see Fig.~\ref{Fig4}). In $Lk=1$, the peak is at $N_W=1$. The chain wrapping the core once is more favorable than wrapping the core twice although the chain loses the adsorption energy. In $Lk=1.25$, there are no large peak; the probability is wide from $N_W=1$ to $N_W=2$. In $Lk=1.5$, $Lk=1.75$, and $Lk=2$, the peak is at $N_W=2$. These results are also same as the result of Fig.~\ref{Fig4}. 

These results indicate that $Lk$ decides the stable wrapping number, how many times the chain wraps the core. In $Lk=0$ to $Lk=1$, the peak shifts from wrapping the core twice to wrapping the core once. Then, in $Lk=1$ to $Lk=2$, the peak shifts from wrapping the core once to wrapping the core twice.
\section{discussions}
\subsection{Internal energy}
\begin{figure}
\includegraphics{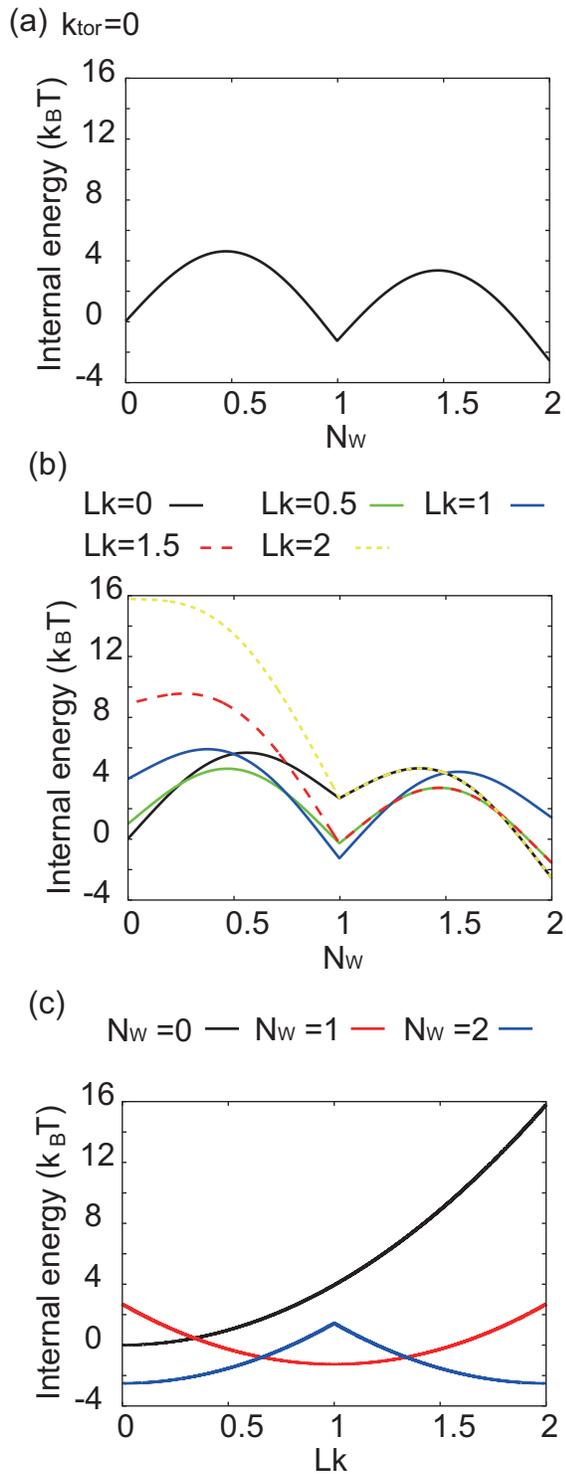}
\caption{The internal energy of wrapping number $N_W$ versus $Lk$ by theoretical results. (a) in the case of $k_{tor} =0$ (without torsional effect). (b) $Lk=0$, $Lk=0.5$, $Lk=1$, $Lk=1.5$, and $Lk=2$ are drawn by black, green, blue, dashed red, and dashed yellow, respectively. (c) The value of internal energy at $N_W=0$ (black), $N_W=1$ (red), and $N_W=2$ (blue) versus $Lk$.}
\label{Fig6}
\end{figure}

A stiff circular DNA molecule is known to exhibit a buckling transition to a supercoiled state upon the increase in $Lk$~\cite{EPL_Guitter}. Here the instability is caused by the balance between the twisting and bending energies. In our case of the twisting of the wrapped complex, the similar competition would be conceivable, but with an important difference that the effective bending modulus can be regarded "negative", i.e., the chain favors the wrapping escorted by the adsorption energy, in the wrapping state (see discussion below). With such a point in mind, we now present a phenomenological theory to describe the properties of the core-chain complex with torsional effect. We refer the internal energy of Ref.~\cite{PRE_Sakaue,PRL_Kulic}. The internal energy of this case can be written as a sum of the adsorption, bending, and torsional energy terms.
\begin{equation}
E(l)=E_{ad}+E_{bend}+E_{tor}
\end{equation}
where $l$ is the chain length around the core. The energetic gain due to the chain adsorption $E_{ad}$ is written as $E_{ad}=\epsilon l$, where $\epsilon$ is adsorption energy density ($\epsilon <0$). The bending energy $E_{bend}$ is written as
\begin{equation}
E_{bend}=\kappa l/2R^2+8\sqrt{\kappa f}(1-\frac{1}{\sqrt 2})| \sin{A}|
\label{bending}
\end{equation}
where $\kappa$ is chain rigidity, $R$ is radius of the core (radius curvature),  $f$ is a extensional force at the end of the chain, and $A$ is described by $A=2\pi l/4\pi R$. 
The second term in eq.~(\ref{bending}), which appears only in the case that both ends of the chain are fixed by pulling them, represents the penalty in the bending energy near the core (see Ref.~\cite{PRL_Kulic}). It has peaks at $l=\pi R$, $3\pi R$, which makes the halfhearted wrapped state with non-integer $N_W$ unfavorable. 
The torsional energy $E_{tor}$ is written as 
\begin{equation}
E_{tor}\approx \int_{0}^{L_C}\frac{k_{tor}}{2} (2\pi)^2(\frac{Tw}{L_C})^2= \frac{k_{tor}}{2} (2\pi)^2 \frac{Tw^2}{L_C}
\end{equation}
where $L_C$ is a contour length of the chain associated with the wrapping. $Wr$ is related to the wrapping number $N_W$. We can write $Tw=Lk-Wr\simeq Lk-N_W$  in the regular wrapping state. $N_W$ is roughly estimated as $N_W=l/2\pi R$. In the case of the antagonistic wrapping state, $1<N_W<2$ and $| Lk| \leq 1$, we use simulational results that the chain wraps the core twice but $Wr=0$; $Wr=| l-4\pi R| /2\pi R$. Finally we get
\begin{equation}
E_{tor}=
\begin{cases}
\frac{2\pi^2k_{tor}}{L_C} (Lk-\frac{l}{2\pi R})^2 &\\
\frac{2\pi^2k_{tor}}{L_C} (Lk-\frac{| l-4\pi R| }{2\pi R})^2 & (1<N_W<2 \mbox{, }| Lk| \leq 1)
\end{cases}
\end{equation}
We set $\epsilon=-5.2$, $R=1.0$, $\kappa=10$, $k_{tor}=10$, $f=0.5$, and $L_C=50$. 
Fig.~\ref{Fig6} shows the internal energy; (a) without torsional effect (b) with torsional effect as a function of $N_W$ and (c) the value of the internal energy at $N_W=0$, 1, and 2 versus $Lk$. In this discussion, we define that the wrapping number $N_W$ increases from 0 to 2 linearly with a increase of $l$ from 0 to $4\pi R$. Without torsional effect (see Fig.~\ref{Fig6} (a)), the most stable is $N_W=2$ and the stable state is $N_W=1$; $E(l)_{N_W=2}<E(l)_{N_W=1}<E(l)_{N_W=0}$. In $Lk=0$ to $Lk=1$, the internal energy at $N_W=2$ increases (see Fig.~\ref{Fig6} (b), (c)). In $Lk=1$ to $Lk=2$, the free energy at $N_W=2$ decreases. The internal energy at $N_W=1$ decreases with $Lk=0$ to $Lk=1$ and increases with $Lk=1$ to $Lk=2$. These results indicate that the chain wrapping the core twice $N_W=2$ is the most stable state in $Lk\simeq 0$ and 2 but not in $Lk\simeq 1$. On the other hand, the chain wrapping the core once $N_W=1$ is the most stable state in $Lk\simeq 1$. Both terms $E_{ad}$ and $E_{bend}$ in eq.~(14) have a linear dependence on $\kappa l/(2R^2) + \epsilon l = \epsilon^* l$, so the wrapping transition takes place when $\epsilon^* < 0  \Leftrightarrow \kappa/(2R^2)<|\epsilon|$ in the absence of the torsional effect. Therefore, the torsional effect, if  included, is a dominant factor in the total free energy of the system. The stable states are decided through decreasing torsional energy. We show a example here. In $Lk\simeq 0$ it is the most stable state to wrap the core twice. When the chain wraps the core twice, the adsorptional energy gain is high and the torsional energy is low because $|Tw|\simeq 0$ ($Lk=0$, $Tw\simeq 0$, and $Wr\simeq 0$). When the chain wraps the core once, the adsorptional energy gain is low and the torsional energy is high because $|Tw|\simeq 1$ ($Lk=0$, $Tw\pm 1$, and $Wr\mp1$).
In $Lk=0$ to $Lk=1$, the stable state shifts from wrapping the core twice to wrapping the core once. Then, in $Lk=1$ to $Lk=2$, the stable state shifts from wrapping the core once to wrapping the core twice. These discussions are consistent with the results of Fig.~\ref{Fig5}. 

We discuss the internal energy here because the entropy of this case should be almost zero: the free energy of this system is estimated as the internal energy. In opposite to this case, fluctuations and entropy are important factors in the case of polynucleosome.
\subsection{force responses}
\begin{figure*}
\includegraphics{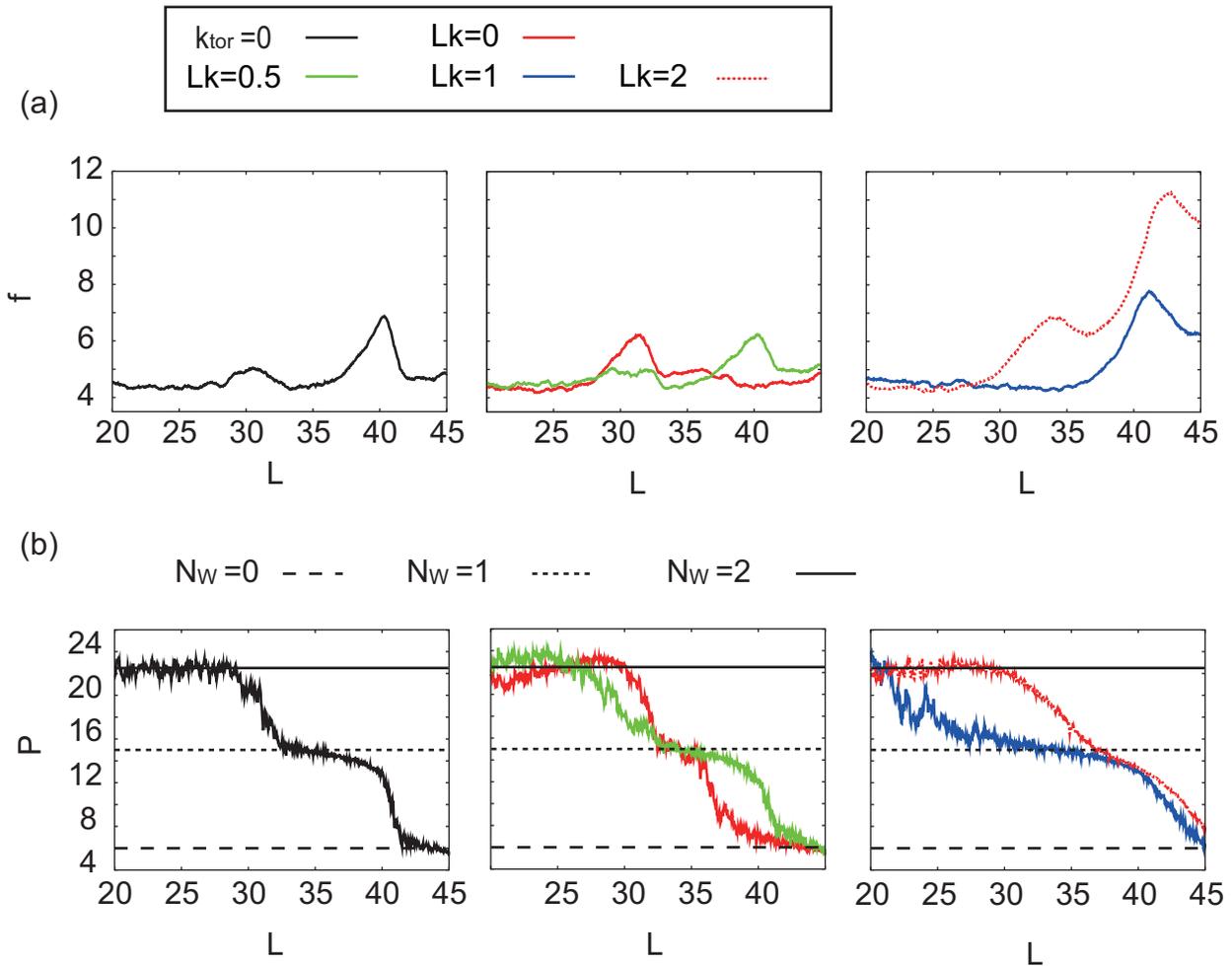}
\caption{Unwrapping process by stretching. The data are averages of ten processes. (a) Extensional force versus the distance of $L$ (end to end distance); $k_{tor} =0$ (without torsional effect), $Lk=0$, $Lk=0.5$, $Lk=1$, and $Lk=2$ are drawn by black, red, green, blue, and dashed red respectively. (b) $N_W$ versus $L$. The wrapping numbers  $N_W=2$ (solid line), 1 (dotted line), 0 (dashed line) are defined by $P=22.5$, 15, 6 respectively.}
\label{Fig7}
\end{figure*}

We study the stability of the wrapping around the core against the stretching. Fig.~\ref{Fig7} (a) is tensional forces of pulled end (at ${\bf r_5}$) versus $L=| {\bf r}_5-{\bf r}_{N-4}| $ (like end-to-end distance). Fig.~\ref{Fig7} (b) shows the number of monomers around the core $P$ versus $L$ with and without torsional effect. In $k_{tor} =0$ (without torsional effect), there are two peaks at $L\simeq 30$, 40; it is a same result in Ref.~\cite{PRE_Sakaue}. At $L\simeq 30$, the chain unwraps the core twice to once; $N_W=2$ changes to 1. At $L\simeq 40$, the chain unwraps the core once to zero; $N_W$ changes to 0. In $Lk=0$, there is a large extensional force at $l\simeq 30$ because the chain wrapping the core twice is stable; $E(l)_{N_W=2}<E(l)_{N_W=1}$. On the other hand, there is a small peak at $l\simeq 37$ because the potential barrier between $E(l)_{N_W=1}$ and $E(l)_{N_W=0}$ is small. The chain easily unwraps the core. In $Lk=0.5$, there are two peaks at $L\simeq 30$, 40. $N_W$ is almost same behavior as the case of $k_{tor} =0$; free energy are $E(l)_{N_W=2}<E(l)_{N_W=1}<E(l)_{N_W=0}$. In $Lk=1$, there is a large extensional force at $l\simeq 37$ because the chain wrapping the core once is stable. There is no peak at unwrapping the core twice to once because the chain unwraps the core easily: the potential barrier between $E(l)_{N_W=2}$ and $E(l)_{N_W=1}$ is small and the chain wrapping core twice is metastable ($E(l)_{N_W=2}> E(l)_{N_W=1} <E(l)_{N_W=0}$). In $Lk=2$, there is a large extensional force at $L\simeq 34$ because the chain wrapping the core twice is stable. These is a very large extensional force at $L\simeq 42$ because the chain not wrapping the core is much more unstable than the chain wrapping the core once; $E(l)_{N_W=2}< E(l)_{N_W=1} <E(l)_{N_W=0}$. In the case of $|Lk|>0$, $E(l)_{N_W=0}$ is large, because $Nw=0$ implies $Wr\simeq 0$ in the present condition, thus the free energy quadratically increases with $Lk$; $E(l)\simeq E_{tor}\simeq (Tw)^2 \simeq (Lk)^2$ (see Fig.~\ref{Fig6} (c)).
It is noted that although the shape of real histone core is close to a cylinder, our spherical core model should capture the essential features in the wrapping-unwrapping transition.

In this manipulation, we fixed five monomers of both ends in order to prevent topological breaking. This corresponds to the case of adsorbing DNA end onto a tip of AFM, a micro bead, and the surface of a glass. In opposite to manipulate one monomer (to fix DNA at one point), it needs stronger force to pull monomers (to fix DNA at several points). There is a kink between a free monomer and a fixed monomer. It causes extra bending energy, which is estimated about $4k_BT$ in this calculation.
\section{Conclusions}
In this study, we have used a model inspired by nucleosome; a single semiflexible chain wraps a spherical core. The stable wrapping number, how many times the chain wraps the core, was studied under topological constraints. With twisting once (quasi-static process), a increase in $Lk$ (topological number) from 0 to 1, the stable states are shifted from the wrapping around the core twice to once. With twisting once more, a increase in $Lk$ from 1 to 2, the stable states are shifted from the wrapping around the core once to twice. Internal energy of the system including the torsional rigidity shows the same results as simulational results. The torsional energy is dominant in the case that the bending energy and the adsorption energy are erased each other; they linearly depends on the length around the core. It restrain the wrapping conformation of the chain. The wrapping number, which is related to $Wr$, and the torsional stress, which is related to $Tw$, are coupled as $Lk=Wr+Tw$. In order to decrease torsional stress, the wrapping number needs change. The stability of the wrapping number is dependent on $Lk$ through torsional stress.

Finally, we mention the unwrapping process by stretching. In the case increasing torsional stress after unwrapping, there is a large extensional force because the chain changes to less stable states. On the other hand, there is a small peak in the case of dereasing torsional stress after unwrapping because the chain easily unwraps the core; the chain is metastable state and the potential barrier is small. These results are easily verified by experiments. Recent experimental technic has developed and it is possible to control torsion. For example, there is a study to twist and pull DNA~\cite{PRL_Besteman}. We think that it is possible to apply this experiment to pulling nucleosome with torsional constraints. We hope that our study will stimulate further experimental and theoretical development in the mechanical stability of chromatin and its link with biological functions.
\section{Acknowledgments}
This work was supported by Japan Society for the Promotion of Science (JSPS) under a Grant-in-Aid for Creative Scientific Research (Project No. 18GS0421) and a fellowship from JSPS (21-1091). YH would like to thank Prof. H. Schiessel, Prof. H. Nakanishi, Dr. N. Yoshinaga, Dr. Y. Takenaka, and Dr. T. Yanao for their valuable discussions.

\end{document}